\documentclass{elsart3p}

\usepackage{graphicx}

\usepackage{amssymb}

\usepackage{epsfig}
\usepackage{amsmath}
\usepackage{xspace}

\usepackage{ifpdf}

\usepackage[%
  breaklinks=true,%
  colorlinks=true,%
  linkcolor=blue,anchorcolor=blue,%
  citecolor=blue,filecolor=blue,%
  menucolor=blue,pagecolor=blue,%
  urlcolor=blue]{hyperref}

\newcommand{\EXIST}{{\it EXIST}\xspace}
\newcommand{\protoexist}{{\it ProtoEXIST}\xspace}
\newcommand{\pe}[1]{{\it ProtoEXIST{#1}}\xspace}

\newcommand{\sS}[1]{\mbox{$\rm{}^{#1}$}}

\newcommand{\Swift}{\mbox{$Swift$}\xspace}

\begin{document}

\begin{frontmatter}



\title{Building Large Area CZT Imaging Detectors \\
	for a Wide-Field Hard X-ray Telescope -- \it ProtoEXIST1
	\thanksref{cor}}
\thanks[cor]{Send correspondence to J. Hong
	: \\
	E-mail: jaesub@head.cfa.harvard.edu, \\
	See the electronic edition for the color version
	of the figures for clarity.
}


\author{J.~Hong, B.~Allen, J.~Grindlay, N.~Chammas\thanksref{mova}}
\address{Harvard-Smithsonian Center for Astrophysics, Cambridge, MA 02138}
\author{S.~Barthelemy, R.~Baker, N.~Gehrels}
\address{NASA Goddard Space Flight Center, Greenbelt, MD 20771}
\author{K.~E. Nelson, S.~Labov, J.~Collins\thanksref{movb}}
\address{Lawrence Livermore National Laboratory, Livermore, CA 94550}
\author{W.~R.~Cook, R.~McLean, and F.~Harrison}
\address{California Institute of Technology, Pasadena, CA 91125}

\thanks[mova]{Now at G2 System LLC, Boston, MA 02110}
\thanks[movb]{Now at XIA LLC, Hayward, CA 94544}

\begin{abstract}


We have constructed a moderately large area (32 cm$^2$),
fine pixel (2.5 mm pixel, 5 mm thick) CZT imaging detector which constitutes 
the first section of a detector module (256 cm$^2$) developed for 
a balloon-borne wide-field hard X-ray telescope, \pe1.  
\pe1 is a prototype for the High Energy Telescope (HET) in 
the Energetic X-ray imaging Survey Telescope (\EXIST), a next generation
space-borne multi-wavelength telescope.  We have constructed 
a large (nearly gapless) detector plane through a modularization scheme
by tiling of a large number of 2 cm $\times$ 2 cm CZT crystals. Our 
innovative packaging method is ideal for many applications such as 
coded-aperture imaging, where a large, continuous detector 
plane is desirable for the optimal performance.  Currently we have been
able to achieve an energy resolution of 3.2 keV (FWHM)
at 59.6 keV on average, which is exceptional
considering the moderate pixel size and the number of detectors
in simultaneous operation.  We expect to complete two modules
(512 cm\sS{2}) within the next few months as more CZT becomes available.
We plan to test the performance of these detectors in a near space 
environment in a series of high altitude balloon flights, the first 
of which is scheduled for Fall 2009.  These detector modules are the first 
in a series of progressively more sophisticated detector units and
packaging schemes planned for \pe2\ \& {\it 3}, which will demonstrate the
technology required for the advanced CZT imaging detectors 
(0.6 mm pixel, 4.5 m\sS{2} area)
required in \EXIST/HET.
\end{abstract}

\end{frontmatter}

\section{Introduction}


The Burst Alert Telescope (BAT), launched aboard \Swift\ - a
multiwavelength Gamma-ray Burst mission, opened a new era of hard X-ray imaging
with the first large array of
CZT detectors in space \cite{Gehrels04}.  CZT detectors are ideal for
use in high energy astronomy as the high Z composition stop X-ray
photons efficiently, the large band gap allows room temperature
operations with excellent spectral resolutions.  Arrangement of
contacts in pixels or strips on these detectors allow
for the precise reconstruction of the positions of incident X-rays. In
addition to X-ray astronomy CZT detectors also have a wide range of
applications including medical imaging and radiologic interdiction.

For our purposes we intend to utilize this technology in
a next generation hard X-ray telescope - the Energy X-ray Imaging Survey
Telescope (\EXIST)\footnote{\url{http://exist.gsfc.nasa.gov}}.  \EXIST is
the leading candidate for the Black Hole Finder Probe under
NASA's Beyond Einstein
Program\footnote{\url{http://universe.nasa.gov/}}, which is currently undergoing further 
development as part of the Advanced Mission Concept Study
Program\footnote{\url{http://www.nasa.gov/home/hqnews/2008/feb/} \url{HQ_08054_Astro_Concept_Studies.html}}.  
The main goals of \EXIST are to
perform an unbiased survey of black holes on all scales and to probe 
the early universe through the detection of high 
redshift ($z\gtrsim7$) GRB's.

The High Energy Telescope (HET) on \EXIST employs CZT detectors in a
coded-aperture imaging telescope, which are similar to the ones in the
\Swift/BAT but significantly advanced.  In coded-aperture imaging, the
sensitivity of the telescope is proportional to the square root of the
detector area.  Therefore, a large area of fine pixel CZT detectors is
essential, in order to achieve the ambitious science goals of the \EXIST
mission.  The construction of such detectors, which can operate using
the limited resources available in a space environment (e.g. power),
is a key challenge for the \EXIST mission.

The CZT detectors on the \Swift/BAT cover about 0.5 m\sS{2} with approximately
31000 $4\times4\times2$ mm$^3$ CZT crystals, where each
crystal works as an individual pixel element, and are 
sensitive between 15 and 150 keV.  \EXIST/HET requires a detector plane
spanning approximately 4.5 m$^2$ and a pixel size of 0.6 mm.  This is
about 10 times the detector area with 7 times finer pixelation than
found in BAT; The HET will have about a factor of 400 more pixels (13M). 
In addition, the HET will cover a wide energy range (5 - 600 keV). The
CZT crystals will 
be 5 mm thick in order to efficiently detect X-rays at energies above
200 keV.  In order to achieve the desired 5 keV threshold extremely
low noise electronics are also required.  The power restrictions for a 
space-based mission also require that the power consumption for each 
channel be extremely low.

These constraints can only be achieved with a very efficient packaging
scheme utilizing sophisticated detector and electronics arrays.
For \EXIST, we plan to employ 13000 closely tiled multi-pixel CZT
crystals ($20 \times 20 \times 5$ mm\sS{3}, 1024 pixels), each of which
is bonded onto an Application Specific Integrated Circuit (ASIC)
that consumes approximately $20 \mu$W/pixel.

Currently we are developing the CZT technology required for
\EXIST in a series of balloon experiments dubbed \protoexist.  In \pe1,
the first in the series, we will construct two large
modules (2 $\times$ 256 cm\sS{2}) of tightly packaged CZT detectors with
2.5 mm pixel. In \pe2, we will refine the backend electronics
to allow for detectors with 0.6 mm pixel, and finally in
\pe3, we will assemble a CZT detector module (256 cm\sS{2})
that can be used in \EXIST.

We have assembled a part ($8\times4$ cm\sS{2}) of the first module for
\pe1 which consists of 8 tiled $2\times2$ cm\sS{2} CZT detectors.
This module is fully operational and, as we continue to acquire CZT crystals,
we expect to complete the two detector modules within the next 3 months
in order to assemble two complete coded-aperture telescopes for the first 
\pe1 balloon flight set for the Fall of 2009.


In this paper we introduce the packaging scheme that enables gap-free
tiling of a large number of CZT detectors (\S\ref{s:module}). We review
unique features in the data collected from these CZT detectors
(\S\ref{s:analysis}) and present the performance for the first
\pe1 sub-module using an \sS{241}Am radioactive
source (\S\ref{s:results}).
Finally we outline our development plan for the next series of 
detector modules for \pe2 \& {\it 3} (\S\ref{s:discussion}).

\section{Modularization Scheme} \label{s:module}

\setcounter{figure}{0}

\begin{figure*}[t] \begin{center}
\includegraphics*[width=\textwidth]{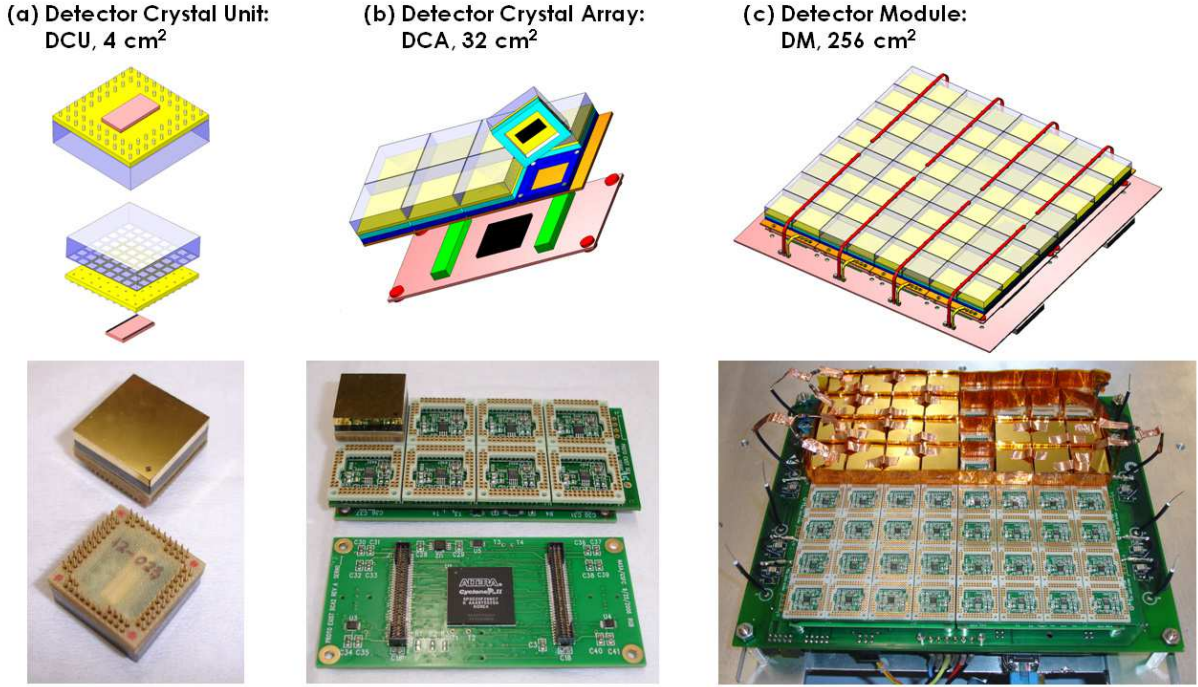}
\caption{The modularization scheme for large area CZT imaging detectors.  (a)
Detector Crystal Unit (DCU): a CZT crystal (blue), an IPB (yellow) and a 
RadNet ASIC (pink) (b) Detector Crystal Array (DCA): 2x4 DCUs, a DCU 
Socket board and a DCA FPGA
board (c) Detector Module (DM): 2x4 DCAs and an FPGA controller board.
In the picture, (a) a red dot on the Au cathode surface of the crystal
indicates the orientation of the crystal with respect to the IPB, and the
RadNet ASIC is partially visible through the protection cover (see also Fig.~\ref{fshield})
(b) one DCU is mounted on the DCA board, (c) 20 DCUs and 8 DCA
boards are mounted in the picture, with each DCU surrounded by a shield
to reduce the interference between the DCUs.  Copper tape is used
to apply the HV bias for easy testing, mounting, and unmounting of the
DCUs.
\label{fpg}}
\end{center}
\end{figure*}

Here we review the detector packaging and operation scheme
developed for \pe1 (Fig.~\ref{fpg}).
An extensive description of the early development of the 
telescope and detector plane can be found in \cite{Hong05,Hong06,Hong07}


\subsection{Detector Crystal Unit (DCU)} 

The basic building blocks from which the \pe1 detector plane is 
constructed are called Detector Crystal Units (DCU).
Each DCU consists of a 5mm thick, $2\times 2$ cm\sS{2} CZT crystal and an
electronic circuit board called the interposer board (IPB) which matches the 
form factor of the CZT.  The interposer contains a single ASIC and is directly
bonded to the CZT on the side opposite the ASIC as shown in Fig.~\ref{fpg}a. 

\subsubsection{CZT crystal} 
We utilize CZT crystals obtained from Redlen 
Technologies\footnote{\url{http://www.redlen.com}} which has developed
a traveling heater CZT growth technique that enables the production of
highly uniform CZT crystals at low cost.  Each Redlen crystal comes
with gold contacts deposited on both the cathode and anode sides.
The anode side has a pattern of 8$\times$8 square pixels with a 2.46 mm
pitch and a 2.0 mm pad size.  These typically exhibit leakage currents
of $\lesssim $ 0.5 nA/pixel under a HV bias of -600 V on the cathode.
The reversal of the anode and cathode or the polarity of the HV bias
results in leakage currents which are approximately an order of magnitude
larger, i.e. there is a preferential orientation for low leakage currents
in Redlen CZT crystals.

Before the attachment of a CZT crystal to an IPB a 
leakage current measurement is performed in order to determine the 
correct orientation of the cathode and anode.  The original Redlen
contacts are then stripped from the CZT and replaced in preparation 
for the IPB bonding process.  Next the CZT
crystals are then bonded onto the IPB's using a low temperature 
($\sim$100--120$^\circ$C) solder bonding technique,
developed at Aguila Technologies\footnote{\url{http://www.aguilatech.com}}, 
which is significantly
less expensive than other bonding techniques that are current available.
The solder bonds provide both the mechanical support and electric
connection between the 64 anode pixels and the matching input pads on
the IPB.  

\subsubsection{Interposer Board (IPB)} 

The IPB is an eight layer board which maps the 2-D 8$\times$8 array of 
anode pixels to 1$\times$64 input pads of the ASIC affixed to the back side
of the IPB.
The design of the IPB has been revised a number of times in order to 
achieve optimal performance. For instance, in order
to reduce the capacitance noise on the ASIC input, we introduced narrow
traces (3 mil) and the low dielectric constant material (Arlon NT55 or ST55)
for the latest revision of these boards.  The capacitance-induced noise
estimates and the measured electronics noise of the bare DCUs (IPB+ASIC,
no crystal) match very well \cite{Hong05, Hong06}, 
demonstrating one of the dominant components of electronics noise we observe
is the large capacitance of the long traces in the IPB's between the 
anode pixels and the ASIC inputs (see also Fig.~\ref{fdcu} in \S5).

\subsubsection{RadNet ASIC}
\begin{figure}[t] \begin{center}
\includegraphics*[width=0.48\textwidth]{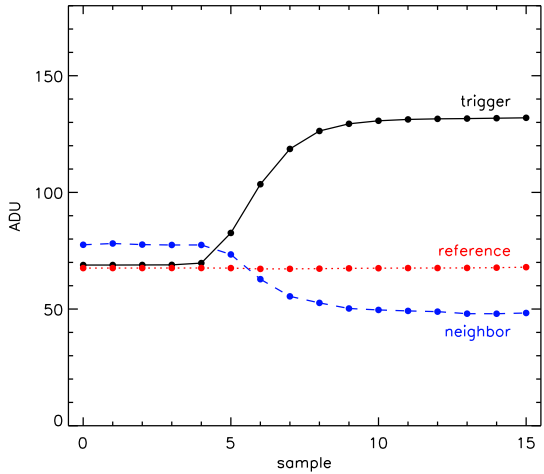}
\caption{Example pulse profiles averaged over events. For each event,
the RadNet ASIC can record 16 samples of the pulse profile for
triggered pixels (solid black), neighboring pixels (dashed blue), 
and the rest of the pixels unrelated to the event (dotted red).
\label{fpp}}
\end{center}
\end{figure}

To process signals from the \pe1 CZT detectors, we employ the RadNet
ASIC. The RadNet ASIC was originally developed for Homeland Security
programs \cite{Craig}, and has inherited the circuitry design from an
ASIC used for a balloon-borne focusing hard X-ray telescope, HEFT
\cite{Cook98}. HEFT was a pathfinder for the upcoming hard X-ray
focusing space telescope,
NuSTAR\footnote{\url{http://www.nustar.caltech.edu/}}, which will
employ CZT detectors at the focal plane.  

The RadNet ASIC has many features that are particularly
attractive for our application. Compared to other known ASICs, one of the
main advantages is its relatively low power consumption
(less than 100 $\mu$W/pixel) which is essential for the operation of a large number
of channels or pixels as required in \EXIST which will contain more than 10 M pixels
with a total power consumption under about 300 W.

In addition, the RadNet ASIC provides a wide dynamic range (10 keV --
1 MeV) with a low intrinsic electronics noise, which is approximately 170
$e^{-}$ root mean square (rms) or about 1.5 keV full width half maximum (FWHM).  
It is also capable of multi-pixel pulse-profile readout, which can provide 
the depth ($z$) information of the interaction as well as $x, y$ and
the energy deposit ($E$). 
The flexible readout mode also allows for the complete
reconstruction of multi-pixel trigger events arising from charge 
splitting or Compton scattering \cite{Hong06,Hong08}.  The 
reconstruction of Compton scattered events makes possible the polarization
measurement of detected X-rays and the full recovery of the total energy 
deposited through multiple interactions within the detector.  
The interaction depth ($z$) is also useful for reducing background
even for events with a single triggered pixel arising from
photoelectric interactions (e.g. low energy events occurring near the
anode side are likely background). It is also important for refining 
imaging with crystals thicker than the pixel width.

\begin{figure}[t] \begin{center}
\includegraphics*[width=0.48\textwidth]{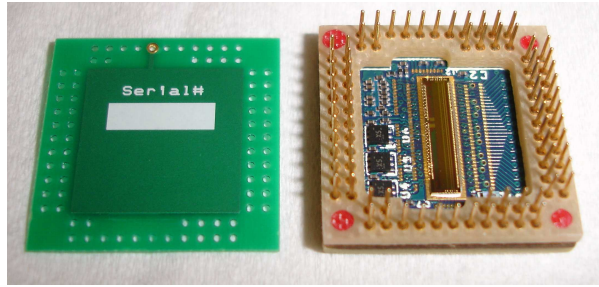}
\caption{The bottom DCU shield and the backside of the bare DCU without a CZT
crystal. The bottom DCU shield fits on the IPB socket and protects
the wire bond of the ASIC and electrically shields the ASIC.
\label{fshield}}
\end{center}
\end{figure}

The basic architecture of the RadNet ASIC can be found in the
literature  \cite{Cook98}; here we briefly review its operation.
Every pixel or channel in the ASIC continuously amplifies input signals,
which are sampled at a user-selected rate from 2 (normal) to 4 MHz and stored in 
a bank of 16 capacitors.  When a trigger occurs, the ASIC continues to sample and
store 8 more values, so that one can extract 8 pre-
and post-trigger samples for the event.
Fig.~\ref{fpp} shows example pulse profiles. The signal height is
determined by the difference between the average of the first 5
pre-trigger sampling points and the average of the last 7 post-trigger
sampling points.  Excluding the rising or falling section of the pulse
profile an averaging over the maximum number of samples possible for
the pre- and post-trigger samples reduces the statistical fluctuation
of electronics noise, which could otherwise dominate the uncertainty in
the pulse height measurement.


Note that the apparent rise time of the pulse profile (about 2 $\mu$sec,
4 or 5 samplings at 2 MHz) does not represent the charge drift time
($<0.5\ \mu$s) in the CZT crystal under the HV but rather is
dominated by charge coupling between the successive sampling
capacitors\footnote{This defect is fixed in the new ASIC for \pe2. See
\S5.}.  Therefore, one cannot use the apparent rise time for deriving
the depth of the interaction, however there exists an alternative method.

\begin{figure*}[t] \begin{center}
\includegraphics*[width=\textwidth]{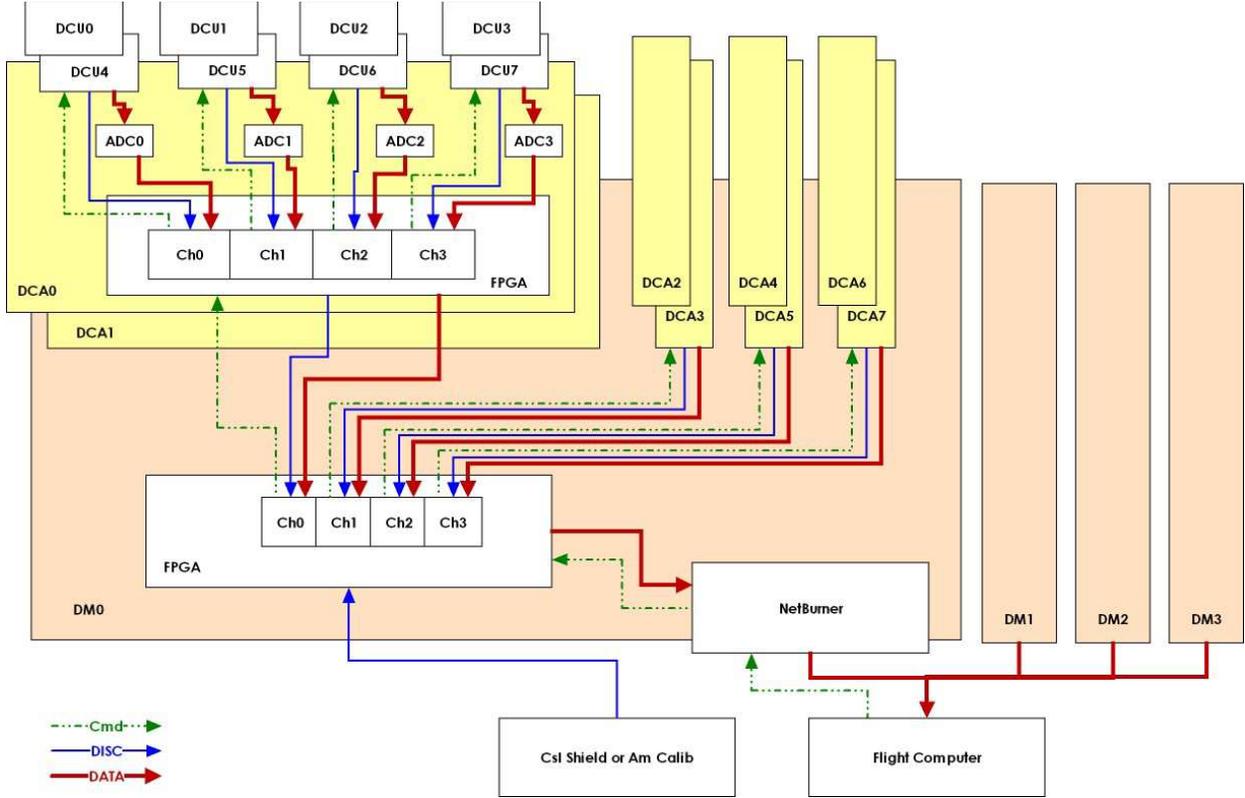}
\caption{Schematics of the data stream and signal processing of \pe1.
The DCA FPGA has four independent channels of the data streams and each
channel processes signals from two DCUs. Similarly the FCB FPGA has
four independent data streams and two DCAs share one stream.
\label{fsch}}
\end{center}
\end{figure*}

The pulse profile readout scheme makes possible the measurement of 
negatively induced charges without need for additional circuitry 
to process signals of the opposite polarity.  As shown in the blue
profile of Fig.~\ref{fpp}, when the post-trigger samples are lower 
in amplitude than
the pre-trigger samples, negatively induced charges are then recorded.
Negative charges are usually induced in pixels neighboring triggered
pixels due to slow-moving holes, and the total sum of the induced signal
in neighbor pixels is proportional to the depth
of the interaction \cite{Eskin99,Hong06,Hong08}.

We have demonstrated this new depth sensing method which takes advantage
of negatively induced charges for the first time in a pixellated
CZT detector \cite{Hong06}. We have achieved sub-mm
depth sensing precision in the 5 mm thick CZT crystal of the first
revision (rev1) DCU.  In a subsequent paper, we will model in detail this
depth sensing procedure and address its prospect for the current version
(rev2) and the future revision (0.6mm pixel) of CZT detectors based on
electrostatic models and measurements \cite{Hong08}.

\subsubsection{DCU shields}
The initial test runs with the simultaneous operation of multiple DCUs
revealed the presence of interference between the DCUs.
This interference and the subsequent increase of the electronics noise are
more prominent in pixels tied to long traces in the IPB under a HV bias.
In order to minimize this we have devised simple electric
shields surrounding the five sides of the IPB in each DCU.  The bottom
side is covered by a simple electronic board with a single ground plane
that is coupled to the ground pin of the DCU through a single trace
(Fig.~\ref{fshield}).  
The bottom shield replaces the partially transparent cover (shown in
Fig.~\ref{fpg}a, the back cover labelled 12-023) that protects the ASIC
wirebonds.  For the four side walls, we have manually
constructed thin shields, which are a thin copper tape sandwiched by
Kapton tape for electrical insulation. The side shields cover all side surfaces
of the DCU from the top of the cathode to the bottom of the rear shield
(Fig.~\ref{fpg}c).  These shields reduce the noise down to the level 
observed when operating a single DCU.



\subsection{Detector Crystal Array (DCA)} 
The next largest element in our modularization scheme is the Detector Crystal 
Array (DCA), which combines $2\times4$ array of DCU's on two 
vertically-stacked electronics boards (Fig.~\ref{fpg}b).  The vertical 
stacking is introduced to package the necessary electronics without
introducing any additional footprint so that we can tile crystals with
a minimal gap for the construction of a large continuous detector plane.

Mounted on the upper board is a 2$\times$8 array of DCU sockets.
In the current revision, we only allow a 400 $\mu$m
gap in between 20 mm x 20 mm crystals for easy (un)mounting DCUs and
the side shields situated between them.  The upper board is about 5 mm longer
than the lower board, extending beyond the edge of the DCU sockets on
one side. This extension is for direct connection between the
HV ground and the ASIC analog ground, so that a series of RC filters
for the HV bias can be electrically and physically 
close to the ASIC for noise 
reduction.  This extension is introduced mainly for
convenience in the initial experiment; the immediate connection can
be still achieved with very minimal or no extension of the board 
beyond the detector plane 16$\times$16 cm\sS{2}.

An Altera\footnote{http://www.altera.com} Cyclone II FPGA (EP2C20F256C7)
is mounted on the lower DCA board to process signals from the DCU's 
by commanding eight ASIC's and four 16-bit ADC's (AD7685BRM) mounted 
on the upper board.  The signals from the DCU's are processed in pairs, so 
there are four data streams on a DCA.  The DCA FPGA is programmed 
to have four corresponding channels, each of which controls two DCU's and 
shares an ADC in parallel in order to minimize the dead time caused by 
the event processing in other pairs.

The signal readout from the RadNet ASIC is highly customizable.  For a
given event, one can read out the 16 samples of the pulse profile from
any number of pixels.  All 16 samples of the 64 pixels for an 
event in a DCU may even be readout for a single event. Such a readout mode provides
a complete snapshot of the ASIC for the event at the expense of
increased signal processing time.  Since a complete snapshot of
the DCU is not always necessary, we have implemented multiple readout 
modes in each of four channels in the DCA FPGA, where
one can read out just one designated pixel (Debug Mode), triggered pixels
(Fast Mode), triggered + neighboring + reference pixels (Normal Mode), or
all 128 pixels of both DCU's in the same pair (Calibration Mode).
Table 1 summarizes the readout modes for a DCU pair.  The main bottleneck
in the event processing pipeline is the ASIC readout which requires about 30 $\mu$s 
for each sample (about 0.5 ms/pixel). This is due in part to the conservative
ASIC readout scheme we have implemented in the DCA FPGA.  In principle
we can increase the data readout speed from the RadNet ASIC by a factor 
of 2 over the current implementation, however the current count rate limit is
well above the 1 cps cm$^{-2}$ that is expected in flight.


\begin{table}[b]
\small
\caption{Readout mode ($N$ trigger pixels per DCU pair)}
\begin{tabular*}{0.48\textwidth}{l@{\extracolsep{\fill}}rrrr}
\hline\hline
Mode		&pixels 		& words		& dead time    & max count \\
		&to read		& /event	& /event       & rate/DCU pair \\
		&			& 		& (msec)       & (cps)\\
\hline
debug		&	1 		& 39		& 0.5 		& 700 \\
trigger		&	$N$		& 23+16$N$	& 0.5 $N$ 	& 700/$N$\\
normal		&	5 .. 128 	& 103 .. 2071 	& 2.5 .. 32 	& 10 .. 160 \\
		&	9$N$+2 		& 55+144$N$	& 4.5 $N$ + 1  	& 350/(4.5$N$+1)\\
calibration	&	128		& 2071 		& 32 	      	& 10\\
\hline
\end{tabular*}
DCU pair: 8 cm\sS{2}
\label{tmode}
\end{table}

\subsection{Detector Module (DM)}
A Detector Module (DM) for a telescope in \pe1 is composed of a
2$\times$4 array of DCA's mounted on a motherboard, called the 
FPGA controller board (FCB), as shown in Fig.~\ref{fpg}c.  
The FCB contains eight sets of sockets for mounting eight DCAs on the top 
side. Another type of Altera Cyclone II FPGA (EP2C20F484C7) is mounted on the
back side for the control and processing of signals from the DCA's.  
The DCA mounting sockets on the FCB's are arranged so that the 20.40 mm pitch 
of the DCU sockets remains constant over the entire DM. Similar to the 
DCA FPGA, we configured the FCB FPGA to have four independent
data channels with each channel reserved to process the data 
stream from two DCA's as illustrated in Fig.~\ref{fsch}.

Two EMCO\footnote{\url{http://www.emcohighvoltages.com}} miniature HV power
supplies (PS) are also attached to the FCB in order to bias the cathode 
surface of 64 crystals normally at --600 V.
Each HVPS provides the bias for half of the DCU's present on the FCB.
The output from the HVPS is split into 4 separately filtered leads that
provide the HV bias for all DCU's on a single DCA.  The lead for each
DCA provides the bias for
8 individual DCU's whose cathodes are coupled using conductive Copper tape.

During the upcoming balloon flight each detector plane will be surrounded 
by four passive side shields and a rear active shield. The passive shield 
is composed of multilayer sheets of Pb/Ta/Sn/Cu, and the active shield 
is a 26$\times$26$\times$2 cm\sS{3} CsI scintillator outfitted with two 
photo-multipliers (PMTs).  For on board calibration of each DM during the 
flight, we employ a small plastic scintillator doped with \sS{241}Am.  
For each 59.6 keV X-ray photon generated in the \sS{241}Am, the associated
$\alpha$-particle triggers the PMT coupled to the scintillator.  We will use 
two \sS{241}Am calibration sources to uniformly irradiate the DM.  The trigger
signals from these shield and calibration PMTs are fed into the DM through the
TTL line for tagging of the shield and calibration source events.  For time
tagging, 1 pulse per second (PPS) TTL pulse from a Global Positioning Unit
(GPS) is fed to the DM.  The detail about the
CsI shield, the onboard \sS{241}Am radioactive sources and the GPS
driven clock will be given in \cite{Allen09}.

When a trigger occurs, the fast trigger signal ($<$ 0.5 $\mu$sec)
alerts the corresponding channel of both the DCA FPGA and the FCB FPGA.
While the DCA FPGA starts the data readout sequence from the ASIC, the
FCB FPGA checks the shield signals, the \sS{241}Am calibration source status,
and the clock counter and then insert them into the data stream when
the DCA FPGA passes the data to the FCB FPGA.


\subsection{Flight Computer and Data Collection}
The DM communicates with the flight computer through ethernet via a
Netburner card\footnote{\url{http://www.netburner.com/}} mounted on the
backside of the FCB for the transfer of data and command and control.
The Netburner card collects data from the four data channels of the FCB 
FPGA and the housekeeping (HK) data of the FCB and passes them on 
to the flight computer. The HK data includes the event trigger rates, the 
\sS{241}Am calibration source rate, the CsI rear shield count rates,
input voltages and currents of the low voltage PS units, temperature
and pressure of the vessel, etc. The Netburner card also controls the
HV setting and passes the commands for the ASICs such as the threshold
setting for each DCU. Note all 64 pixels in each DCU share a common
threshold.


%



\begin{figure}[t] \begin{center}
\includegraphics*[width=0.40\textwidth]{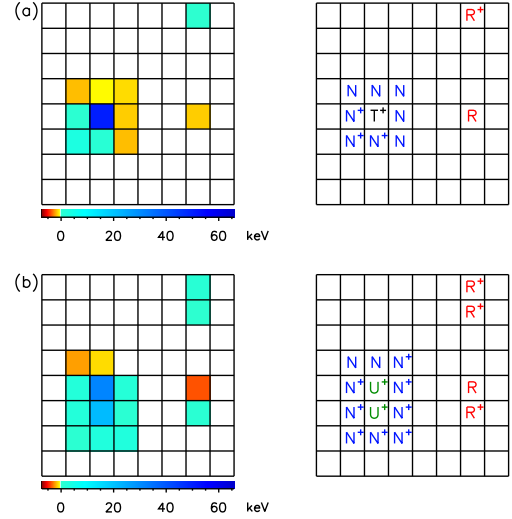}
\caption{Example events in a DCU; the top two panels show
the calibrated signals and the pixels marked for readout from
a single pixel trigger event, and the bottom two panels show
the same for a two pixel trigger event. The (black) 'T's are
for isolated trigger pixels, the (green) 'U's for unisolated
trigger pixels, the (blue) 'N's for neighbor pixels, and
the (red) 'R's for reference pixels. The '+'s indicate positive
energy deposits. 
\label{fexa}}
\end{center}
\end{figure}

\section{Data Analysis} \label{s:analysis}
\begin{figure*}[t] \begin{center}
\includegraphics*[width=\textwidth]{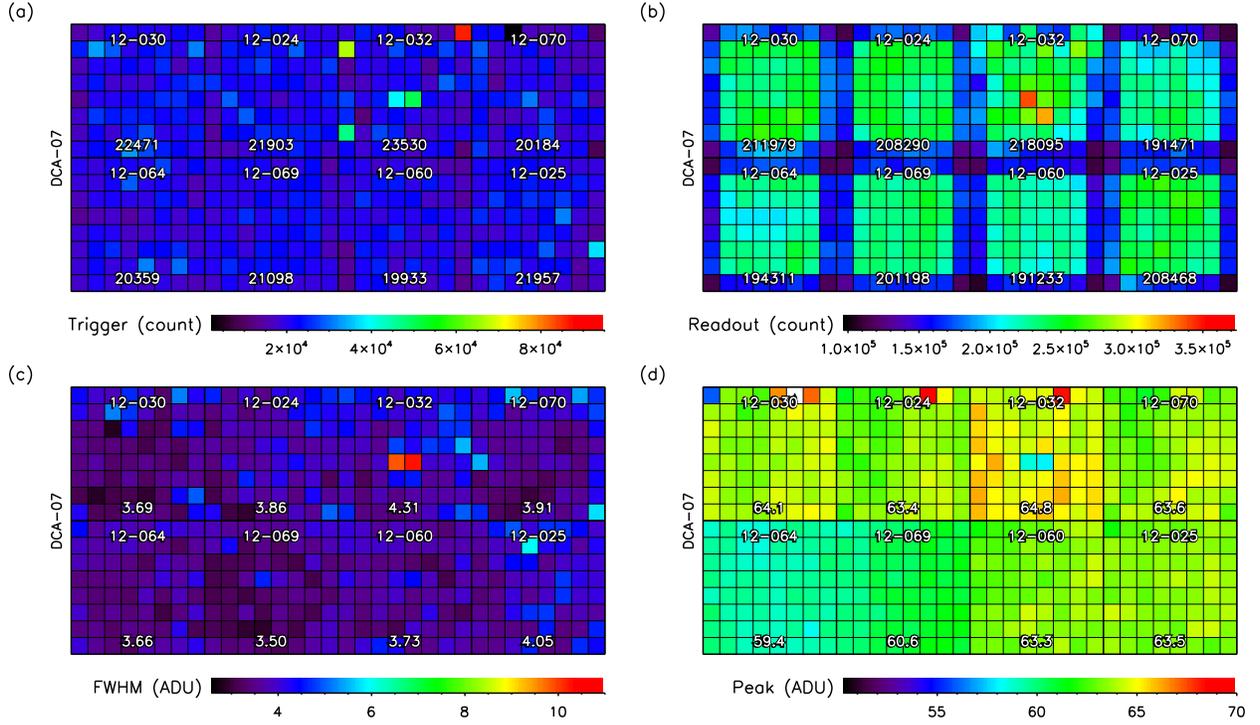}
\caption{Raw data taken from an \sS{241} Am source: (a) the trigger count
map, (b) the readout count map, (c) the energy resolution (FWHM in
ADU) at 59.6 keV,  and (d) the peak position (in ADU) of the pulse height
histogram for 59.6 keV X-rays. 1 keV is roughly 1.05 ADU.
\label{fraw}}
\end{center}
\end{figure*}

As time of this writing, we have 22 working DCUs. Here we present the
analysis results of the data taken with the first eight DCUs using
a \sS{241}Am radioactive source shining on the cathode side of the 
detectors.

The data were taken in the normal readout mode, which reads out all 
triggered pixels plus all pixels adjacent to triggered pixels on the same
DCU and 2 unrelated reference pixels for zero point calibration.  
Fig.~\ref{fexa} shows an example of the charge deposit map and the
readout pixels for an event with an isolated trigger pixel (a total of
11 pixels to be read out) and the same for a charge split event with two
trigger pixels (16 pixels to be read out).
In principle, one can read out the signals from the neighbor pixels in the
neighbor crystals when a trigger occurs on edge or corner pixels, but for
now we limit the neighbor pixel readout within one DCU because the side
DCU shield prevents the neighbor crystal from inducing negative charges.

Due to the complexity and the unique nature of the data provided by the
\pe1 CZT detectors, it is worthwhile to explore the properties of the
raw data and the calibration procedure for future reference.


\subsection{Raw Data and Calibration}
Fig.~\ref{fraw} shows pixel map views of the raw \sS{241}Am data: (a) the
total trigger count, (b) the readout count map, (c) the energy resolution
(FWHM) and (d) the peak position of the pulse height histogram for the
59.6 keV X-rays from the source.  Each crystal is labelled by the DCU ID
(e.g.~12-030) and the pixel-averaged quantity of the display (e.g.~22471
trigger counts for 12-030).  The raw pulse height is given in our custom
unit - ADU, where 1.05 ADU is approximately 1.0 keV.

The trigger count map shows there are a few hot pixels. The readout 
map appears as a tiled pattern of squares which is due to the dearth of
neighboring pixels on the edges, which when triggered would otherwise 
initiate a read sequence.
The energy resolution ($2.23 \sigma$) and the peak values ($E_p$)
are derived from a gaussian fit with no constant term ($A_0 \exp[-(E -
E_p)^2/4/\sigma^2]$) to the pulse height histogram of each pixel.

\begin{figure*}[t] \begin{center}
\includegraphics*[width=\textwidth]{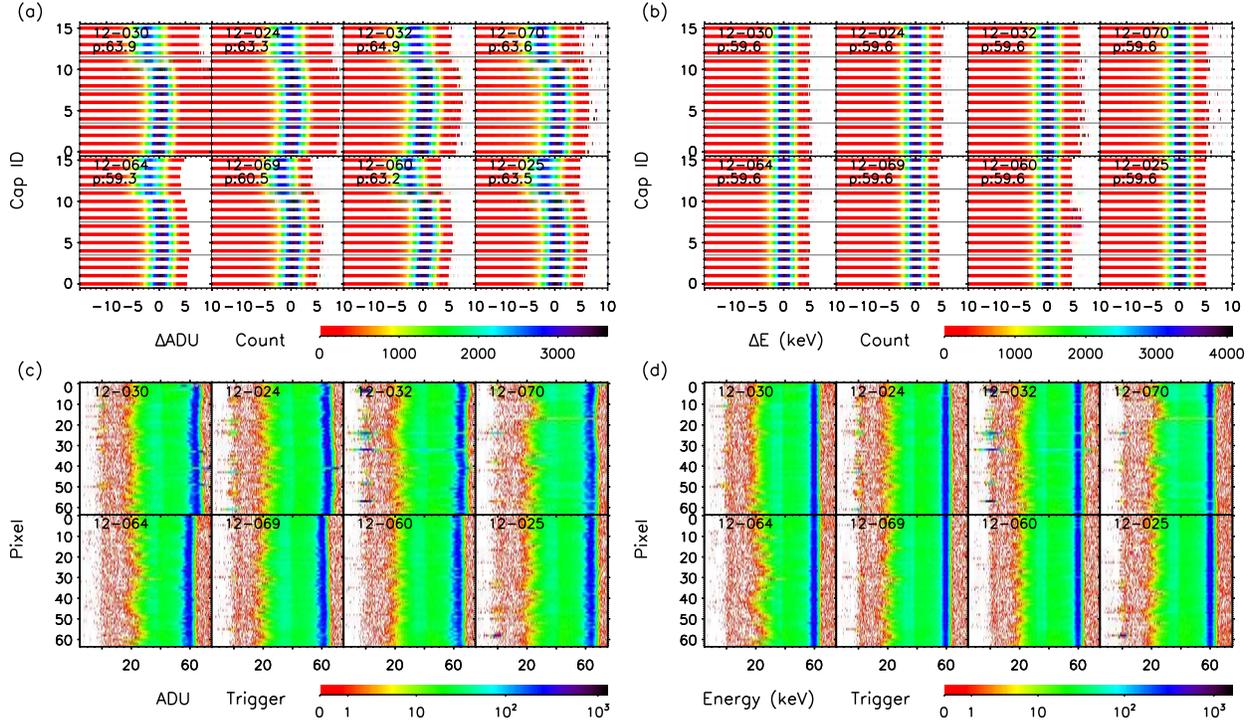}
\caption{The CapID and pixel dependence of the pulse height: (a) and (b)
CapID vs.~the pulse height relative to the peak position, (c) and (d)
pixel ID vs.~the pulse height. (a) and (c) are in the raw pulse
height (ADU), and (b) and (d) are in the calibrated energy (keV).
\label{fcap}}
\end{center}
\end{figure*}

As previously mentioned, the pulse height is calculated by subtracting
the average of the first 5 pre-trigger samples from the average of last 7
post-trigger samplings.  A trigger can occur during samples by any one
of the 16 capacitors and small variations in the capacitance values of 
the 16 capacitors may cause the resulting pulse height to vary depending on
the triggering capacitor ID (CapID).  Therefore a CapID dependent calibration
must be applied in order to properly convert the raw ADU values into energies.
Fig.~\ref{fcap}a shows the CapID dependence of the pulse height in the
DCU stacked data. The pulse height ($x$-axis) is given in the relative
ADU from the pulse peak position.  This dependence also
varies from pixel to pixel (Fig.~\ref{fcap}c), so that for calibration,
the pulse peak of each pixel for each CapID has to be derived.
The right panels in the Fig.~\ref{fcap} show the result after
calibration which has removed the pixel- and CapID-driven variation.

\subsection{Calibrated Data}
\begin{figure*}[t] \begin{center}
\includegraphics*[width=\textwidth]{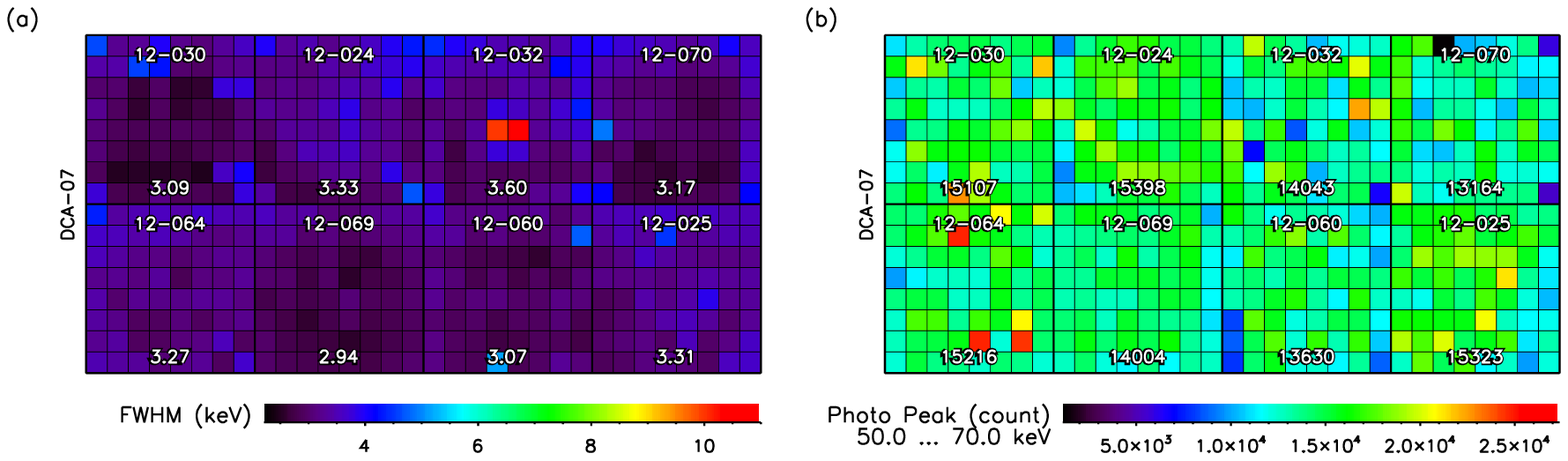}
\includegraphics*[width=\textwidth]{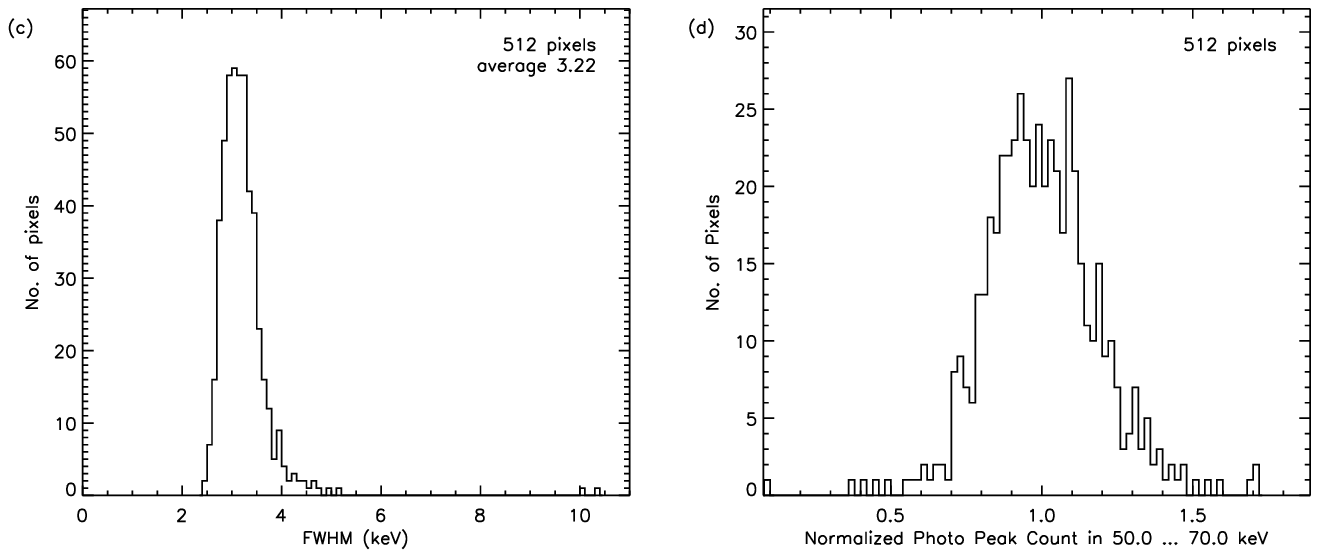}
\caption{Calibrated data: (a) the energy resolution map (FWHM),
(b) photo-peak count map (50--70 keV), (c) the pixel distribution
of the energy resolution, and (d) the same of photo-peak counts.
\label{fcal}}
\end{center}
\end{figure*}

Fig.~\ref{fcal} shows the energy resolution (FWHM) and the photo peak
count map (counts in 50--70 keV) for the calibrated data of
the isolated triggered pixels. The figure also shows
the pixel distributions of the energy resolution
(FWHM) and the photo peak count. The average 
FWHM is 3.2 keV.

The two pixels in the DCU 12-032 still show a relatively large FWHM ($\sim
10$ keV) even after the calibration, and we suspect cross talk between
the two pixels is caused 
by a malformed bond due to their physical proximity. 
Additionally the electric pulser data before bonding the crystal did not 
show any unusual behavior in these two pixels.

The photo-peak count map or its pixel distribution shows a noticeable
variation ($\sim$ 20\% variation for 68\% of the pixels) of measured
counts across all pixels despite the relatively uniform irradiation of the
DCUs. Note we expect about 1\% variations from the statistical fluctuation
under the assumption of a Poisson distribution and the measured number
of counts.  This large variation could be partially the result of high
count rates on the DCU's, which were about a factor of 3 higher than the
maximum rate the current readout mode can handle without any event loss.
Further investigation is required to identify the cause of this large
variation. We also plan to measure the absolute quantum
efficiency of the CZT detectors using the on board \sS{241}Am calibration 
sources, which allows the measurement of the absolute number of X-ray photons
emitted by the source as well as the number of captured X-rays in the 
CZT detectors.

\section{Results} \label{s:results}

\begin{figure*}[t] \begin{center}
\includegraphics*[width=\textwidth]{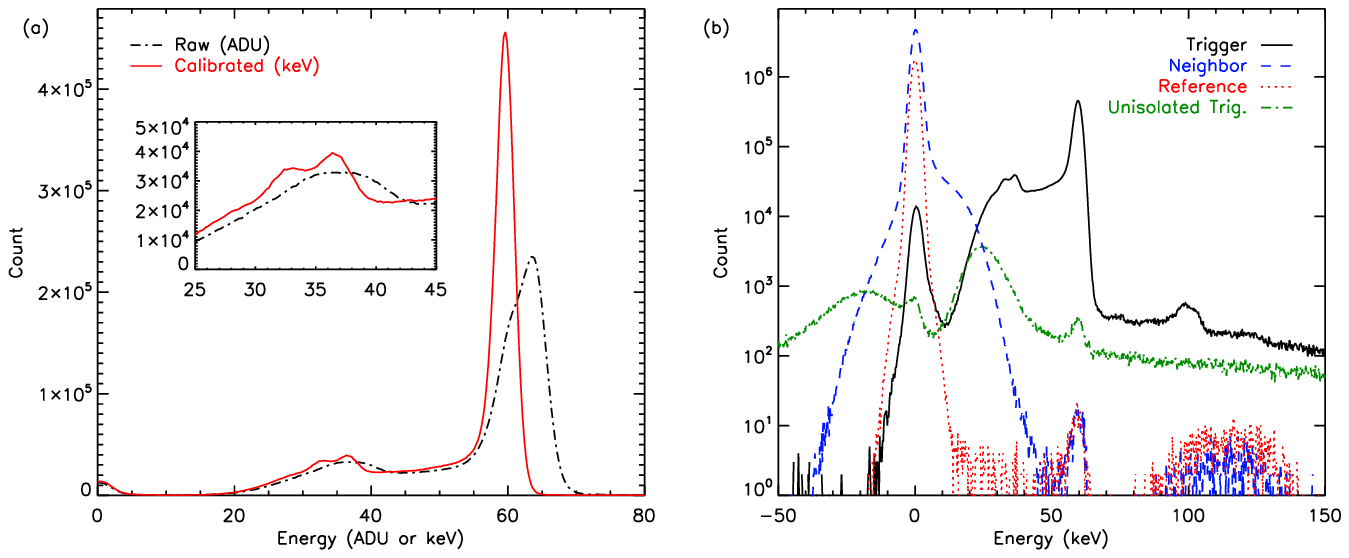}
\caption{The energy histogram:
(a) the pulse height histogram of the raw data (dashed-dot black) and
the calibrated energy histogram (solid red) from the 8 DCUs, and (b)
the calibrated energy histograms of four types of charge
collection in pixels. The calibrated histogram in (a) shows 3.2 keV
FWHM at 59.6 keV and two well resolved escape peaks (the inset).  The
energy histograms in (b) show charge collection for the isolated
triggered pixels (solid black), non-triggering
neighbor pixels (dashed blue), non-triggering reference pixels (dotted red), and
unisolated trigger pixels (dashed-dot green).  \label{fsp} }
\end{center}
\end{figure*}

Fig.~\ref{fsp}a shows the calibrated energy histogram of all 8 DCUs
(red) in comparison with the raw pulse height histogram (black).
For these histograms, we are using the data from isolated triggered pixels
('T's in Fig.\ref{fexa}, see below).  The calibrated histograms show an
energy resolution of 3.2 keV (FWHM) at 59.6 keV and the two escape peaks
around 33 and 37 keV (the red line in the inset in Fig.~\ref{fsp}) are
well resolved.  

Fig.~\ref{fsp}b compares the calibrated energy histograms of four
different cases of charge collection in pixels: isolated triggered pixels
with no triggered neighbors (black 'T's), untriggered neighbor pixels
(blue 'N's), untriggered reference pixels (red 'R's), and unisolated
triggered pixels with triggered neighbors (green 'U's).
Fig.~\ref{fsp}b and the right panels in the Fig.~\ref{fexa} use
the same color or code scheme for easy comparison.
All four histograms show a mild pile-up at 60, 100,
and 120 keV respectively.

The isolated triggered pixels (black in Fig.~\ref{fsp}b) manifest themselves in events where all or most of
the energy deposited from the  interaction is confined to a single pixel and the
induced charges in neighbor pixels are below the trigger threshold.
The histogram falls off just below the 30 keV due
to the threshold set around 25 to 30 keV. A peak around $E= 0$ keV in 
the histogram is due to a few hot pixels, which can be also spotted in 
Fig.~\ref{fcap}d (e.g.~DCU 12-032).  These hot pixels are suspected to be 
due to the unusually high capacitance value in one of 16 sampling capacitors in the
ASIC and a relatively high base line for the pixel in question. Since there 
is no real induced signal in these pixels, the resulting pulse height is 
centered around zero.

The pulse height histogram of the untriggered neighboring pixels shows
a large shoulder on the both sides of the peak at 0 keV (blue in
Fig.~\ref{fsp}b). The negative side is due to the negative charges induced
by slow-moving holes and the positive side is dominated by the
charge split from the triggered pixels, where the energy deposited in
the untriggered pixel does not exceed the threshold.  The histogram
for the reference pixels (red in Fig.~\ref{fsp}b) is relatively symmetric 
with respect to 0
keV and does not have large shoulders as expected since no
energy is deposited into these pixels or their neighbors.  Note the signal
of a reference pixel is used for calibrating the deposited charge of the
same pixel in other events, but not for the other trigger or neighbor
pixels of the same event.

The energy histogram of the unisolated triggered pixels (green in
Fig.~\ref{fsp}b) shows a peak around 30 keV, which is due to the fact
that it is about half of the energy of the 59.6 keV X-ray photons 
from the source and the threshold is set around the same 30 keV.  
For instance, if the charge splits into two
pixels with 20 and 40 keV each, only the energy deposit of 40 keV will
generate an isolated trigger and the 20 keV deposit will be readout in
the untriggered neighbor pixels.  On the other hand, if the charge
splits into 30 and 30 keV each, both pixels will likely trigger resulting in
an unisolated trigger (e.g.~Fig.~\ref{fexa}b).

\begin{figure}[t] \begin{center}
\includegraphics*[width=0.40\textwidth]{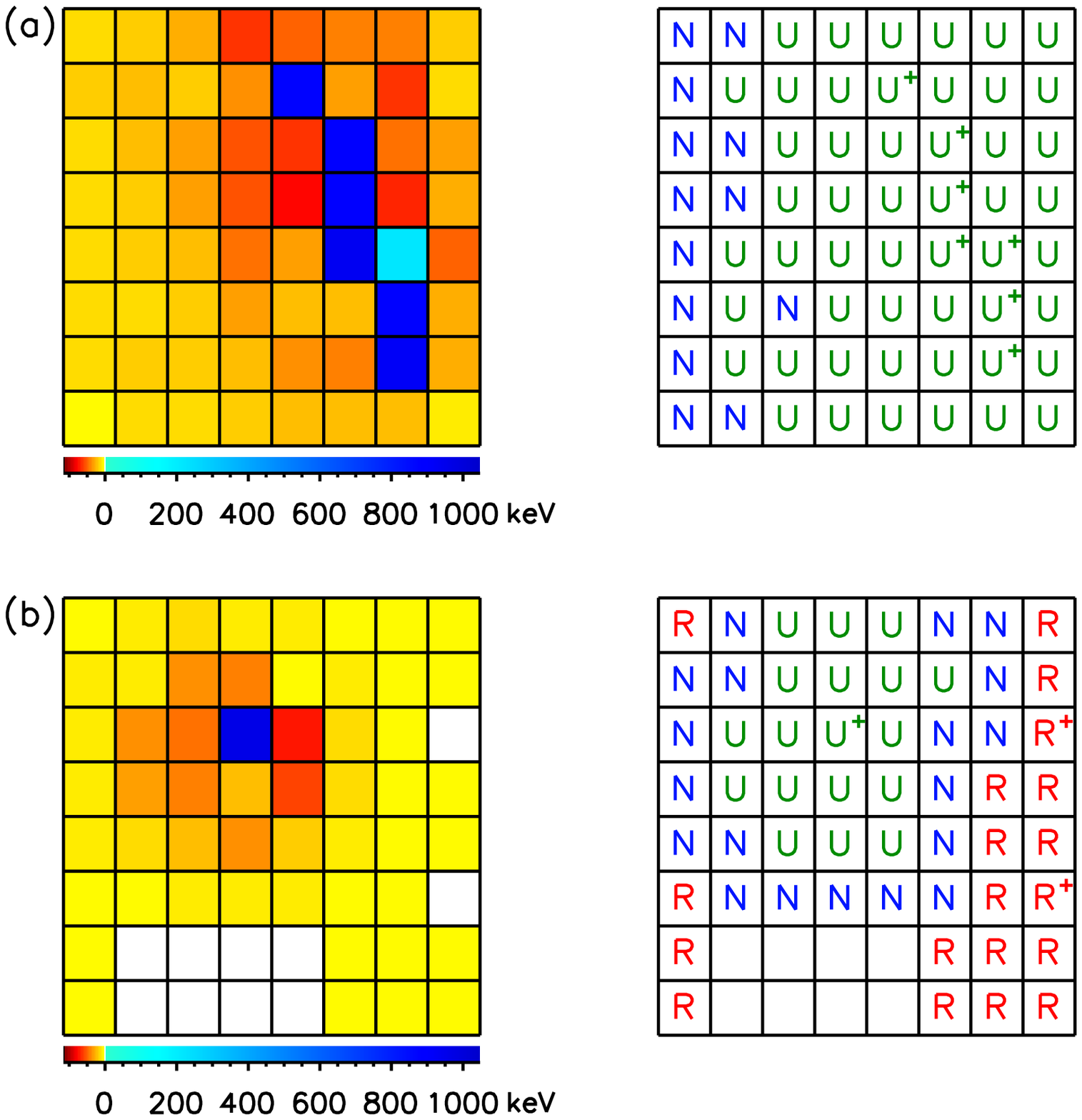}
\caption{Example events triggered by muons in a DCU; 
the top panels show an event with muon coming at an angle
and the bottom panels show a case of near normal incidence.
The symbols in the right panels are the same as in Fig.~\ref{fexa}.
\label{fexa2}}
\end{center}
\end{figure}

The energy histogram of the unisolated triggered pixel indicates some
events can trigger a pixel with an induced signal below threshold or
even with a negative signal.  This is usually due to high energy X-ray
events or charged particle (mostly muons) induced events initiating a
large electron cloud.  In these events, soon after the interaction, 
the large electron cloud causes multiple pixels to induce positive 
charges large enough to trigger simultaneously.
As the electron cloud drifts toward a few anode pixels immediately
underneath the origin of the cloud, the induced
charges in the neighbor pixels begin to be dominated by the slow-moving
holes, and as a result, these neighboring pixels can record a sub-threshold
signal or even negative signals with their own trigger.
Fig.~\ref{fexa2} shows the energy deposit and the trigger pixels of
such events.  A more detailed analysis will be shown elsewhere
\cite{Hong08}.

%

%

%
%

\section{Discussion} \label{s:discussion}


Table~\ref{noise} summarizes the electronics noise and the energy
resolution of the \pe1 CZT detectors and the estimated projections
for future CZT detectors for \pe2 \& {\it 3} and \EXIST.  The 
performance of the rev1 CZT detectors is taken from the measurement
of one good rev1 DCU (01-050) among several DCUs we have produced 
\cite{Hong06}.  For the rev2 CZT detectors, we use the results of 
the 8 DCUs shown in the previous section. The italicized numbers
are the estimated values, based on simple projections based on comparison of
the results from $^{241}$Am tests.

For the rev2 CZT detectors, we have a 60--70\% yield for
bare DCUs (ASIC + IPB) with all 64 working pixels and 2.1--2.5
keV FWHM average electronics noise; the 30--40\% loss
is due to faulty ASIC's.  After bonding the carefully
selected and processed CZT crystals to the bare DCU's, we have a 90\%
yield for completed DCU's with the 3.1--3.5 keV FWHM at 60 keV.  This is
a drastic improvement in yield over the rev1 DCU's.  For \pe1
CZT detectors, we have achieved our goal for energy resolution ($\lesssim$
3.5 keV FWHM at 59.6 keV) with relatively high yield. The current
energy resolution implies $<$1\% energy resolution (FWHM) at 662 keV.



\begin{table*}[b]
\small
\caption{The electronics noise and energy resolution of \pe\ CZT detectors (FWHM)} 
\begin{tabular*}{0.98\textwidth}{l@{\extracolsep{\fill}}rrrrrr}
\hline\hline
Stage			& \multicolumn{2}{c}{\pe1}	& \multicolumn{2}{c}{\pe1}	& \multicolumn{2}{c}{\pe2, \it 3}	\\
			& \multicolumn{2}{c}{rev1}	& \multicolumn{2}{c}{rev2}	& \multicolumn{2}{c}{\& \EXIST} 	\\
\cline{2-3} \cline{4-5} \cline{6-7}
			& (keV)		& (\%)		& (keV)		& (\%)		& (keV)		& (\%)	\\
\hline
ASIC intrinsic 		&		&		&		&		&0.8\sS{a} 	&	\\
After mounting on the IPB	& $3.1 \pm 0.5$	&		& $2.1\pm 0.3$	&		&		&	\\
After crystal bonding	& $3.7 \pm 0.4$	&		& $2.5\pm 0.2$	&		&1.8\sS{a}	&	\\
under HV (-600V)	& $4.6 \pm 0.6$	&		& $3.0\pm 0.5$	&		&2.0\sS{a}	&	\\
60 keV  (\sS{241}Am)	& $4.4 \pm 1.3$ & 7.3\%		& $3.2\pm 0.5$ 	& 5.3\% 	&2.2\sS{a}	&3.7\%	\\
122 keV (\sS{57}Co)	& $5.2 \pm 1.4$ & 4.3\%		&\it 3.8 	& \it 3.1\% 	&\it 2.6 	&\it 3.0\%	\\
356 keV (\sS{133}Ba)	& $7.3 \pm 1.9$	& 2.0\%		&\it 5.3 	& \it 1.5\% 	&\it 3.7 	&\it 1.0\%	\\
665 keV (\sS{137}Cs)	& $7.9 \pm 2.5$ & 1.2\%		&\it 5.7 	& \it 0.9\% 	&\it 3.9 	&\it 0.6\%	\\
\hline
\end{tabular*}
The italicized numbers are the estimates based on the measurements.
The measurements show the average eletronics noise or energy resolution (FWHM) of the
pixels and the $1\sigma$ distribution.  
The electronics noise is measured by the 
internally generated pulses in the ASICs and the true electronics noise is
estimated to be slightly smaller than shown here due
due to intrinsic variations associated with the internal pulses.
(e.g. the average FWHM of the electronics noise under the HV (4.6 keV)
is smaller than the average FWHM of the energy histogram at 60 keV
(4.4 keV) in the rev1 \pe1 CZT detectors).  (a) The preliminary results by \cite{Cook08}.
\label{noise}
\end{table*}

CZT imaging detectors in \EXIST will require 0.6 mm pixels in order to achieve
high angular resolution.  This high pixel density will be applied for
the \pe2 \& {\it 3} CZT detectors.  In the \pe2 CZT detectors, we will
make use of an existing ASIC called the DB ASIC with 1024 channels.  The DB
ASIC has been developed for the NuSTAR project,
and it inherits and advances the circuitry design from the HEFT ASIC and
RadNet ASIC.  In the DB ASIC the variations in capacitance
values of the 16 sampling capacitors and the coupling between subsequent
capacitors have been substantially reduced.  This improvement allows for a very
low threshold ($<$ 5 keV) and the true representation of the charge drift time
in the pulse profile, and may even eliminate the necessity of the
CapID-driven calibration.  In addition, the DB ASIC also includes the
on-chip ADC, which simplifies the back-end electronics and minimize the
possible interference among the units.

In NuStar, the DB ASIC is directly bonded to the CZT crystal,
eliminating the need for an IPB, which is where the
designation DB (i.e. direct bond) originates from.  The ASIC input form
factor matches a CZT crystal with 32$\times$32 arrays of anode pixel of
0.6 mm pitch. The matching form factor is also of practical necessity
due to the large number of pixels (1024).

\begin{figure}[t] \begin{center}
\includegraphics*[width=0.48\textwidth]{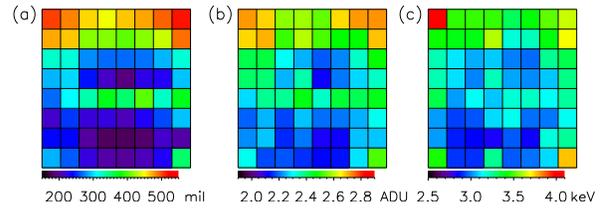}
\caption{Comparison of the trace length and energy resolution: (a) IPB
trace lengths, (b) electronics noise (FWHM in ADU) from the pulser
data before crystal bonding, and (c) energy resolution (FWHM in keV)
at 59.6 keV.  For (b) and (c), we average the results over the 8 DCUs, and
in (c), we exclude two outlier pixels in DCU 12-032.
\label{fdcu}}
\end{center}
\end{figure}

Fig.~\ref{fdcu} compares the trace lengths in the current rev2 IPB (a) with
the electronics noise from the pulser data before crystal bonding (b)
and the energy resolution at 59.6 keV for the 8 DCUs used in \S3 (c).
The electronics noises and the energy resolutions are averaged over the
8 DCUs to reveal the pixel pattern more clearly.  The similarity of the
patterns in the pixel map indicates one of the main noise contributors is
the capacitance noise as a result of long trace lengths.  
Therefore, the direct bonding
approach in the DB ASIC makes possible the energy resolution of less
than 2 keV by removing the IPB (and the long traces in the IPB). 
An additional advantage will be the low leakage 
current for a single pixel which will be about a factor of 16 lower 
than in \pe1 due to the smaller (0.6 mm) pixel size.


In the \pe2 CZT detectors, we will employ this DB ASIC and apply a
packaging scheme similar to that employed in \pe1.
We plan to assemble one DM with 256 cm\sS{2} of the active 
area for \pe2, in order to demonstrate the feasibility of signal
processing in a large array of the high pixel density CZT detectors.

\begin{figure}[t] \begin{center}
\includegraphics*[width=0.48\textwidth]{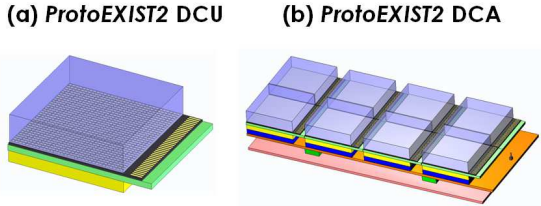}
\caption{Modularization scheme for \pe2 CZT detectors.
\label{fpe2} }
\end{center}
\end{figure}

Fig.~\ref{fpe2} shows our packaging scheme for the DCU and DCA of the \pe2
CZT detectors. In \pe2\ the CZT crystal is directly bonded to the ASIC,
which sits on a small circuit board.  Since the DB ASIC requires about 3
mm of space in one direction for control and power lines in addition to
the 1024 input pads, we need to allow for a 5 mm gap between DCU's in
order to accommodate wirebonds on one side. The DCA board design will
likely be simplified due to the inclusion of an ADC in the DB ASIC.
The energy resolution estimates in the last
column of Table~\ref{noise} are based on the initial measurement of 
the CZT detector unit with the DB ASIC \cite{Cook08}.

For efficient packaging of a large CZT detector for \EXIST, we need
to minimize the gap between detectors. In addition, the power
consumption of the DB ASIC is currently about 100 $\mu$W/pixel 
which is about a factor five higher than the required value for \EXIST.  
Therefore for \pe3 we plan to modify the DB ASIC producing
a lower power version, the `EX' ASIC, to meet these requirements.  
For gapless packaging we are currently considering two possible
options. One method is designed to take advantage of micro-via technology,
which would allow the control lines to be brought out to the backside of 
the chip, such that one could match the foot-print of the ASIC and the 
CZT crystal. In an alternative approach we would either re-package the ASIC
in a smaller form factor (1.5$\times$1.5 cm\sS{2}) or use slightly
bigger CZT crystals (e.g. 2.5$\times$2.5 cm\sS{2} with 0.78 mm pixel) and use an
interposer board with de-magnifying traces of minimal lengths ($<$
2 mm), which map out 32$\times$32 anode pixels to 32$\times$32 ASIC
input pads.  We also plan to assemble one DM with a total active area of
256 cm\sS{2} for \pe3, which will definitively demonstrate the technologies
necessary for the HET on \EXIST.

\section{Summary}
A large contiguous array of CZT detectors has many applications in hard
X-ray imaging. We have demonstrated our innovative packaging scheme for
the construction of large CZT detector planes with relatively small pixels.
The current working module (32 cm\sS{2} with 2.5 mm pixel) shows about 3.2
keV energy resolution at 59.6 keV and we expect
to complete two modules (512 cm\sS{2}) as we acquire more CZT crystals.
Our packaging scheme will be progressively expanded to demonstrate
advanced CZT detectors required for the HET on EXIST (4.5 m$^2$, 0.6 mm 
pixel size, 5-600 keV).

This work is supported in part by NASA APRA grant NNG06WC12G.  Portions
of this work were performed under the auspices of the U.S. Department
of Energy by Lawrence Livermore National Laboratory under Contract
DE-AC52-07NA27344.











\bibliographystyle{elsart-num}

\end{document}